\documentclass[reprint,amsmath,amssymb,superscriptaddress,aps]{revtex4-2}
\usepackage{graphicx}
\usepackage{dcolumn}
\usepackage{bm}
\usepackage[utf8]{inputenc}
\usepackage[english]{babel}
\usepackage[T1]{fontenc}
\usepackage{amsmath}
\usepackage{hyperref}
\usepackage{braket}
\usepackage{bbold}
\usepackage{float}
\usepackage{tabularx}
\usepackage{tikz}
\usepackage{lipsum}

\begin{document}

\title{Universal set of quantum gates for the flip-flop qubit in the presence of 1/f noise}
\author{Elena Ferraro}
 \email{elena.ferraro@mdm.imm.cnr.it}
 \affiliation{CNR-IMM Agrate Unit, Via C. Olivetti 2, 20864 Agrate Brianza (MB), Italy}
 \author{Davide Rei}
\affiliation{Quantum Technology Lab, Dipartimento di Fisica "Aldo Pontremoli", Universit\`a degli Studi di Milano, Via G. Celoria 16, 20133 Milano, Italy}
 \affiliation{CNR-IMM Agrate Unit, Via C. Olivetti 2, 20864 Agrate Brianza (MB), Italy}
  \author{Matteo Paris}
\affiliation{Quantum Technology Lab, Dipartimento di Fisica "Aldo Pontremoli", Universit\`a degli Studi di Milano, Via G. Celoria 16, 20133 Milano, Italy}
\author{Marco De Michielis}
 \email{marco.demichielis@mdm.imm.cnr.it}
\affiliation{CNR-IMM Agrate Unit, Via C. Olivetti 2, 20864 Agrate Brianza (MB), Italy}

\begin{abstract}
Impurities hosted in semiconducting solid matrices represent an extensively studied platform for quantum computing applications. In this scenario, the so-called flip-flop qubit emerges as a convenient choice for scalable implementations in silicon. Flip-flop qubits are realized implanting phosphorous donor in isotopically purified silicon, and encoding the logical states in the donor nuclear spin and in its bound electron. Electrically modulating the hyperfine interaction by applying a vertical electric field causes an Electron Dipole Spin Resonance (EDSR) transition between the states with antiparallel spins $\{\ket{\downarrow\Uparrow},\ket{\uparrow\Downarrow}\}$, that are chosen as the logical states. When two qubits are considered, the dipole-dipole interaction is exploited allowing long-range coupling between them. A universal set of quantum gates for flip-flop qubits is here proposed and the effect of a realistic 1/f noise on the gate fidelity is investigated for the single qubit $R_z(-\frac{\pi}{2})$ and Hadamard gate and for the two-qubit $\sqrt{iSWAP}$ gate.
 \end{abstract}


\maketitle

\section{Introduction}
Quantum computing applications encompass a variety of different scientific, social and economical contexts, from fundamental science to finance, security and medical sectors. In the variegated landscape of physical qubits, semiconducting qubits encoding quantum information in the spin of electrons or nuclei confined through artificial atoms, such as quantum dots and donor atoms, are an established powerful tool \cite{Morton-2011,Laucht-2015,Veldhorst-2017,Vandersypen-2017,Chan-2018,Dzurak-2019}. In particular, donor spins have unprecedented advantages in terms of their long coherence time, high control and scalability. When a phosphorus donor is implanted in silicon, eventually using isotopically purified nanostructures ($^{28}$Si) to drastically reduce magnetic noise, another advantage comes out, that is the integrability with the Complementary Metal-Oxide-Semiconductor (CMOS) technology for the qubit fabrication \cite{maurand2016cmos}.

The main obstacle to the realization of a donor-based quantum processor following Kane's seminal proposal \cite{Kane-1998} is the use a short-range interaction (10-15 nm) among qubits, namely the exchange interaction between the donor bound electrons, that requires a strong near-atomic precision in the donor implantation. One way to get around this issue, relaxing the strict requirement on donor placement, is based on the possibility to access long-range electric dipole-dipole interaction, thus reaching qubit distance up to hundreds of nm. In Ref. \cite{Tosi-2017}, a qubit in which an electric dipole is created sharing the electron between the donor and the interface has been proposed and called \textit{flip-flop} qubit \cite{Tosi-2018,Boross-2016,Simon-2020,Calderon-2021}. This qubit is manipulated by microwave electric field that modulates the hyperfine interaction. In addition, a dc electric field is applied to perform qubit rotations along the $\hat{z}$-axis of the Bloch sphere, and an ac electric field is required to perform $\hat{x}$ and $\hat{y}$ rotations. The electrical control clearly makes the flip-flop qubit more sensitive to charge noise, that typically shows a 1/f spectrum, representing a not negligible source of decoherence \cite{Paladino-2014}.

In this paper, we present a universal set of quantum gates for quantum computation with flip-flop qubits. It is composed by the $R_z(-\frac{\pi}{2})$ and the Hadamard (H) one-qubit gates and the $\sqrt{iSWAP}$ two-qubit gate. It is indeed possible to demonstrate that a universal gate set is $G=\{H,\Lambda(S)\}$, where $\Lambda(S)$ is a two-qubit gate in which the operation S is applied to the target qubit if and only if the control qubit is in the logical state $\ket{1}$, for example the CNOT gate \cite{Preskill-2015}. Moreover, a construction of the CNOT gate using only $R_z(-\frac{\pi}{2})$, H and $\sqrt{iSWAP}$ gates is feasible \cite{Schuch-2003}. For each gate operation, we consider the effect of the charge noise using the 1/f model for the power spectral density.

The paper is organized as follows. In Section II we present the flip-flop qubit, its Hamiltonian model and the study on the noise effects on the fidelity for the single-qubit gates. In Section III we focus on the description of two interacting flip-flop qubits including the dipole-dipole interaction in the Hamiltonian model and then showing a fidelity analysis on the $\sqrt{iSWAP}$ two-qubit gate. Section IV contains the main conclusions.

\section{Flip-flop qubit}
\begin{figure}[H]
\centering
\includegraphics[width=0.8\columnwidth]{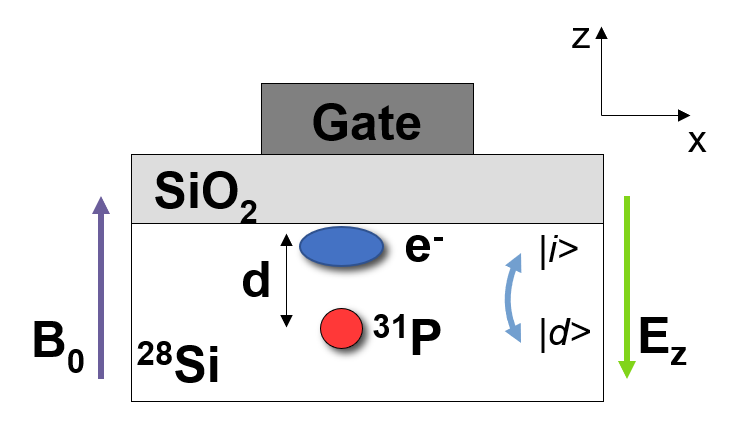}
\caption{Scheme of the flip-flop qubit. A donor atom of $^{31}P$ is embedded in a bulk of $^{28}Si$ at a distance d from the $Si/SiO_2$ interface. The metal gate controls through an electric field $E_z$ the electron position between the nucleus ($\ket{d}$) and the interface ($\ket{i}$) with the dielectric. A constant magnetic field $B_0$ is also applied.}\label{flipflop}
\end{figure}

The flip-flop qubit is realized embedding a phosphorous $^{31}$P donor atom in a $^{28}$Si nanostructure at a depth $d$ from the interface (SiO$_2$ layer) as shown in Fig. \ref{flipflop} 

A vertical electric field $E_z$ applied by a metal gate on top, controls the position of the electronic wavefunction \cite{Tosi-2017,Tosi-2018}. The electronic spin (S=1/2) is described in the basis $\{\ket{\downarrow},\ket{\uparrow}\}$ and has a gyromagnetic ratio $\gamma_e$= 27.97 GHzT$^{-1}$, while for the nuclear spin (I=1/2) the basis is denoted by $\{\ket{\Downarrow},\ket{\Uparrow}\}$ and the gyromagnetic ratio is $\gamma_n$= 17.23 MHzT$^{-1}$, they interact through the hyperfine coupling A. Applying a large static magnetic field $B_0$, (i.e. $(\gamma_e+\gamma_n)B_0\gg A$), the eigenstates of the system are the four qubit states:
$\{\ket{\downarrow\Uparrow},\ket{\downarrow\Downarrow},\ket{\uparrow\Downarrow},\ket{\uparrow\Uparrow}\}$. Electrically modulating the hyperfine interaction A by $E_z$ causes an Electron Dipole Spin Resonance (EDSR) transition between the states with antiparallel spins $\{\ket{\downarrow\Uparrow},\ket{\uparrow\Downarrow}\}$, that are in turn chosen to encode the qubit. 

\subsection{Hamiltonian model}
The flip-flop qubit Hamiltonian model $H^{ff}$ is given by the sum of three contributions \cite{Tosi-2017,Simon-2020}
\begin{equation}\label{Ham}
H^{ff}=H_{orb}+H_{B_0}+H_A.
\end{equation}
The first term is the orbital Hamiltonian that reads (in units of Hz):
\begin{equation}
H_{orb}=-\frac{\varepsilon_0}{2} \sigma_z-\frac{d e E_{ac}(t)}{2 h}\left(\frac{d e \Delta E_z}{h \varepsilon_0}\sigma_z+\frac{V_t}{\varepsilon_0}\sigma_x\right),
\end{equation}
where $V_t$ is the tunnel coupling between the donor and the interface potential wells; $\Delta E_z=E_z-E_z^0$ where $E_z^0$ is the vertical electric field at the ionization point, i.e. the point in which the electron is shared halfway between the donor and the interface; $\varepsilon_0=\sqrt{V_t^2+(d e \Delta E_z/h)^2}$ is the energy difference between the orbital eigenstates, where $h$ is the Planck's constant, $d$ is the distance from the interface, hereafter $d$=15 nm, and $e$ is the elementary charge. For completeness, the ac electric field $E_{ac}(t)$ is also included and is equal to $E_{ac}\cos(\omega_E t+\phi)$. It is applied in resonance with the flip-flop qubit, i.e. $\omega_E=2\pi\epsilon_{ff}$, where $\epsilon_{ff}$ is the flip-flop qubit transition frequency, and $\phi$ is an additional phase. The Pauli matrices are expressed in the basis of the orbital eigenstates: $\sigma_z=\ket{g}\bra{g}-\ket{e}\bra{e}$ and $\sigma_x=\ket{g}\bra{e}+\ket{e}\bra{g}$, where $\ket{g} (\ket{e})$ is the ground (excited) state of the orbital part of the Hamiltonian. We point out that the electron position operators, i.e. $\sigma_z^{id}=\ket{i}\bra{i}-\ket{d}\bra{d}$ and $\sigma_x^{id}=\ket{i}\bra{d}+\ket{d}\bra{i}$, where $\ket{i} (\ket{d})$ denotes the interface (donor) electron position, are expressed in the orbital eigenbasis, by the following relation: $\sigma_z^{id}=\frac{d e \Delta E_z}{h \varepsilon_0}\sigma_z+\frac{V_t}{\varepsilon_0}\sigma_x$ and  $\sigma_x^{id}=-\frac{V_t}{\varepsilon_0}\sigma_z+\frac{d e \Delta E_z}{h \varepsilon_0}\sigma_x$.

The second term in Eq.(\ref{Ham}) is the Zeeman interaction due to the presence of the static magnetic field $B_0$ oriented along the $\hat{z}$ axis and includes also the dependence of the electron Zeeman splitting on its orbital position through the quantity $\Delta_{\gamma}$ (that in the following we set to -0.2\%). The Zeeman term $H_{B_0}$ may be written
\begin{equation}
\begin{split}
H_{B_0}=&\gamma_e B_0\left[\mathbb{1}+\left(\frac{\mathbb{1}}{2}+\frac{d e \Delta E_z}{2 h \varepsilon_0}\sigma_z+\frac{V_t}{2 \varepsilon_0}\sigma_x \right)\Delta_{\gamma}\right]S_z+\\
&-\gamma_n B_0 I_z,
\end{split}
\end{equation}
where $\mathbb{1}$ is the identity operator on the orbital subspace and $B_0$=0.4 T. 

Finally, the hyperfine interaction is given by
\begin{equation}
H_A=A(\Delta E_z)\left(\frac{\mathbb{1}}{2}-\frac{d e \Delta E_z}{2 h \varepsilon_0}\sigma_z-\frac{V_t}{2 \varepsilon_0}\sigma_x\right) \textbf{S}\cdot\textbf{I},
\end{equation}
where the dependence of the hyperfine coupling A by the electric field is highlighted. To obtain the functional form of $A(\Delta E_z)$, that changes from the bulk value $A_0$=117 MHz to 0 when the electron is at the interface, we fit the results from Ref.\cite{Tosi-2017} with the function $A_0/(1+e^{c\Delta E_z})$, obtaining $c=5.174\cdot10^{-4}$ m/V. 

We assume a qubit working temperature of T=100 mK, so as to ensure that the thermal energy $k_BT$ (where $k_B$ is the Blotzmann constant) is always lower than the minimun qubit energy $\epsilon_{ff}=\sqrt{(\gamma_e+\gamma_n)^2B_0^2+A(E_z)^2}$, that is $\simeq$ 11 GHz.

We chose to describe the flip-flop qubit expressing its Hamiltonian in the complete eight-dimensional basis $\{\ket{g\downarrow\Uparrow},\ket{g\downarrow\Downarrow},\ket{e\downarrow\Uparrow},\ket{g\uparrow\Uparrow},\ket{g\uparrow\Downarrow},\ket{e\downarrow\Downarrow},\\\ket{e\uparrow\Uparrow},\ket{e\uparrow\Downarrow}\}$, where the states are ordered from the lower to the higher corresponding energy values, and $\{\ket{g\downarrow\Uparrow},\ket{g\uparrow\Downarrow}\}$ are respectively the $\{\ket{0},\ket{1}\}$ logical states.

\subsection{Single-qubit gates}
In this subsection we present the results obtained analyzing the entanglement fidelity for the single-qubit $R_z(-\frac{\pi}{2})$ and H gates when the 1/f noise model is included \cite{Epstein-2014,Zhang-2017,Yang-2019PRA,Paladino-2014,Ferraro-2020-sr}. The 1/f noise model is based on the definition of the Power Spectral Density (PSD) that is inversely proportional to the frequency and is given by $S(\omega)$ = $\alpha/(\omega t_0)$, where $\alpha$ is the noise amplitude, that does not depend on $\omega$ and $t_0$ is the time unit. Following Ref. \cite{Yang-2016} we generated the 1/f noise in the frequency domain as
\begin{equation}\label{eq:S1overf}
n(\omega)=m(\omega)^{-1/2} e^{i\varphi(\omega)},
\end{equation}

where $m(\omega)$ is generated from a standard Gaussian white process and the phase factor $\varphi(\omega)=[0,2\pi]$ is chosen uniformly. To obtain the noise in the time domain, we calculate the inverse Fourier transform and then multiply the result by the noise amplitude $\alpha$.

\subsubsection{$R_z(-\frac{\pi}{2})$ gate}
The rotation of an angle $-\pi/2$ around the $\hat{z}$-axis of the Bloch sphere is obtained in the following way: a dc electric field $\Delta E_z(t)$ is adiabatically swept to move the electron from the interface at an idling electric field $\Delta E_{idle}$ to the value of clock transition (CT) $\Delta E_{ct}$, and back. The adiabatic set-up consists of a first fast step of duration $\tau_1$, reaching an intermediate value $\Delta E_{int}$, and a second slower step of duration $\tau_2 $ reaching $\Delta E_{ct}$. Then, the electron remains at the CT for a time T before coming back at the idling. The ac electric field is zero. In Table \ref{tab:RotZ_pi} all the parameters set to implement the $R_z(-\frac{\pi}{2})$ gate are reported, $T_{gate}$ denotes the total gate time.
\begin{widetext}
\begin{center}
\begin{table}[htbp!]
\caption{\label{tab:RotZ_pi}$R_z(-\frac{\pi}{2})$ gate parameters}
\begin{center}
\begin{ruledtabular}
\begin{tabular}{ccccccccc}
$\Delta E_{idle}$ & $\Delta E_{int}$ & $\Delta E_{ct}$ & $\tau_1$ & $\tau_2$ & T & $V_t$ & K & $T_{gate}$\\
{[V/m]} &{[V/m]} &{[V/m]} & {[ns]} & {[ns]} & {[ns]} & {[GHz]} &   & {[ns]}\\
\hline
10000 & 500 & 290 & 1.7 & 3.5 & 21.6 & 11.29 & $\simeq$20 & 31.9\\
\end{tabular}
\end{ruledtabular}
\end{center}
\end{table}
\end{center}
\end{widetext}


The coefficient $K$, representing the adiabatic factor, is calculated as the minimum value between the charge adiabatic factor $K_c$ and the spin-orbit adiabatic factor $K_{so}$. Both are derived from a simple two-level Hamiltonian model \cite{Tosi-2017} 
\begin{equation}
H=\Delta\sigma_z+\Omega\sigma_x
\end{equation}
and the adiabatic condition holds when
\begin{equation}
K=\left|\frac{\omega_{eff}}{\dot{\beta}}\right|\gg 1,
\end{equation}
where $\omega_{eff}=\sqrt{\Delta^2+\Omega^2}$ and $\beta=\arctan\left(\frac{\Omega}{\Delta}\right)$. For the $R_z(-\frac{\pi}{2})$ gate, in order to find $K_c$ for the charge qubit, we use $\Delta_c=\frac{\pi e d \Delta E_z}{h}$ and $\Omega_c= \pi V_t$, whereas for the spin-charge coupling we use $\Delta_{so}=\pi\delta_{so}$, where $\delta_{so}=\varepsilon_0-\varepsilon_{ff}$ with $\varepsilon_{ff}=\sqrt{(\gamma_e+\gamma_n)^2B_0^2+A(\Delta E_z)^2}$, that is the flip-flop qubit transition frequency, and $\Omega_{so}=2\pi g_{so}$ where $g_{so}=\frac{A}{4}\frac{V_t}{\varepsilon_0}$.

\begin{widetext}

\begin{figure}[H]
\centering
a)\includegraphics[width=0.8\textwidth]{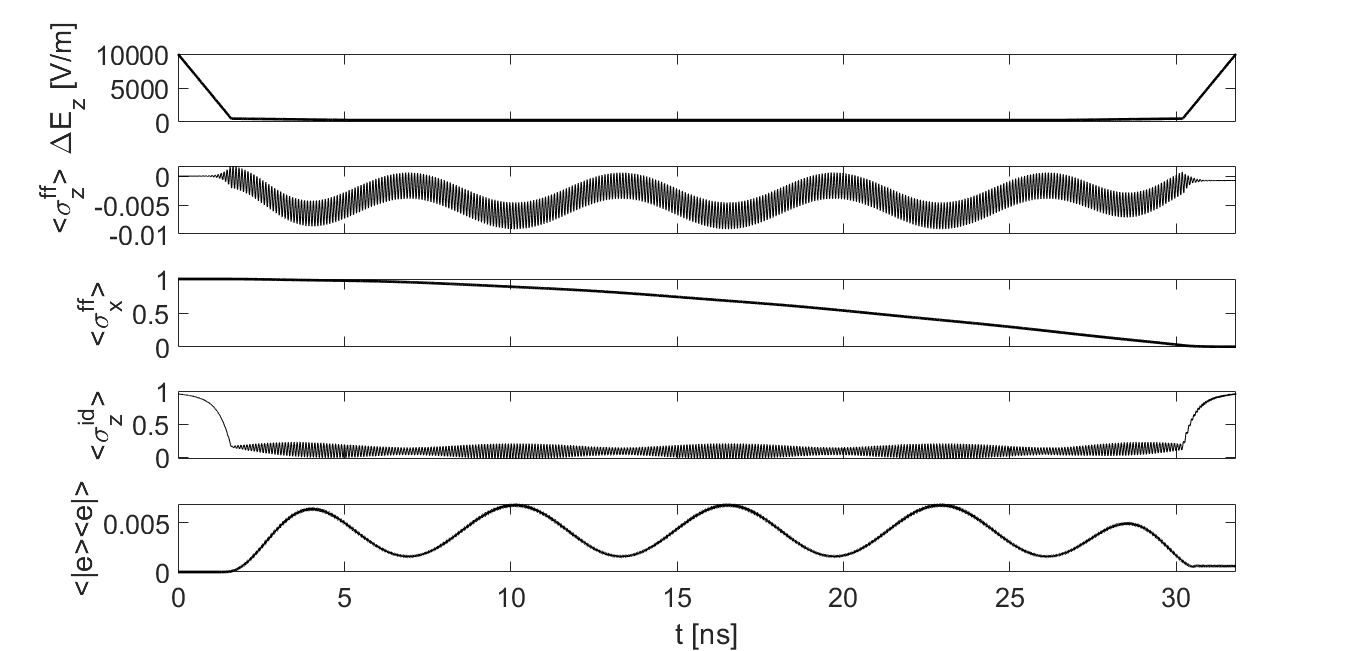}\\
b)\includegraphics[width=0.8\textwidth]{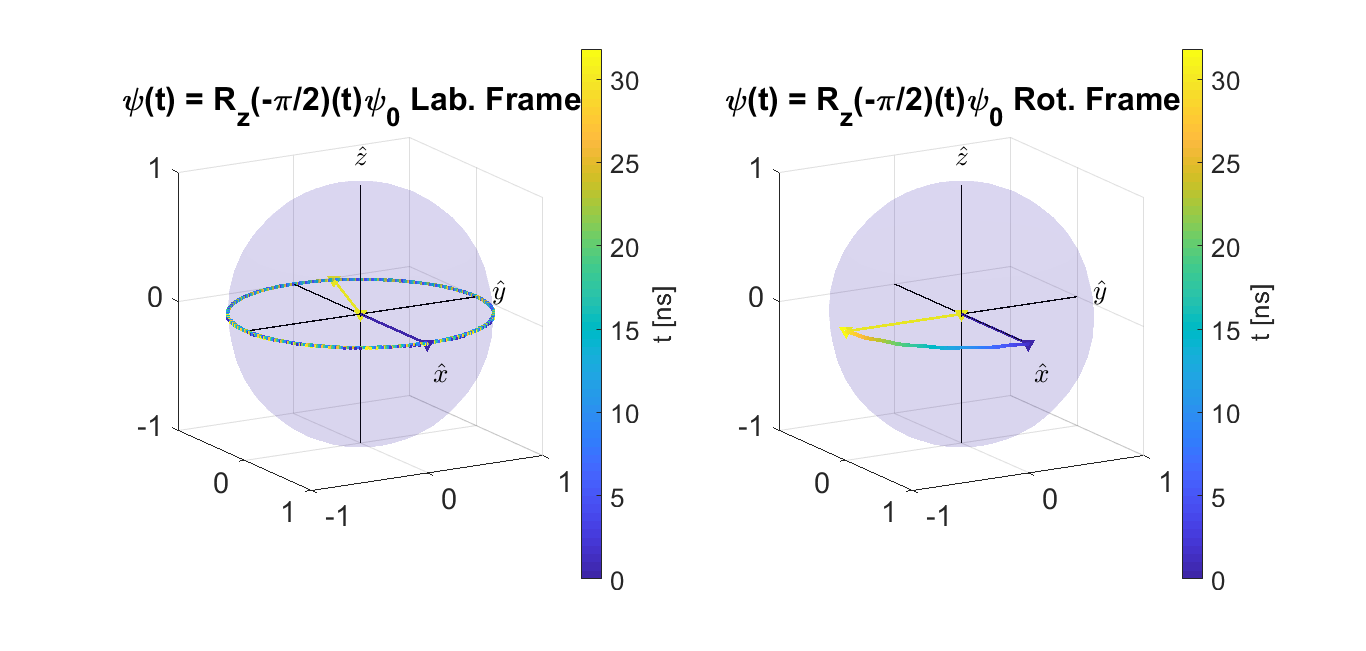}\\
\caption{(a) Time evolution of the dc applied electric field $\Delta E_z(t)$ and of the mean values of the operators $\sigma^{ff}_z$, $\sigma^{ff}_x$, $\sigma^{id}_z$ and $\ket{e}\bra{e}$ during the evolution of the $R_z(-\frac{\pi}{2})$ gate for the initial state $\ket{\psi_0}=\frac{\ket{0}+\ket{1}}{\sqrt{2}}$. (b) Bloch sphere representation of the $R_z(-\frac{\pi}{2})$ gate. The blue (yellow) arrow represents the initial (final) state in the laboratory frame (left) and in the rotating frame (right).}\label{Rz}
\end{figure}

\end{widetext}

In Fig. \ref{Rz}(a) we report the dynamical behaviour of the dc field $\Delta E_z(t)$ as well as the mean values of the single qubit operators: $\sigma^{ff}_z=\ket{\uparrow\Downarrow}\bra{\uparrow\Downarrow}-\ket{\downarrow\Uparrow}\bra{\downarrow\Uparrow}$, $\sigma^{ff}_x=\ket{+_x^{ff}}\bra{+_x^{ff}}-\ket{-_x^{ff}}\bra{-_x^{ff}}$ with $\ket{\pm_x^{ff}}=(\ket{\uparrow\Downarrow}\pm e^{-i2\pi\varepsilon_{ff}^{t=0}}\ket{\downarrow\Uparrow})/\sqrt{2}$, $\sigma^{id}_z$ and the charge excitation $\ket{e}\bra{e}$ in the flip-flop subspace during the evolution of the $R_z(-\frac{\pi}{2})$ gate- To provide an example, we have chosen to start from the initial condition $\ket{\psi_0}=\frac{\ket{0}+\ket{1}}{\sqrt{2}}$. Fig. \ref{Rz}(b) shows the Bloch sphere representation of the $R_z(-\frac{\pi}{2})$ gate operation when the qubit is observed in the laboratory frame (left) and in a frame rotating at the angular frequency of an idling qubit (right). The yellow arrow represents the expected final state obtained after the application of the sequence.

\subsubsection{Hadamard gate}
The Hadamard gate acts on the qubit as a rotation of an angle $\pi$ around the $(\hat{x}+\hat{z})/\sqrt{2}$ axis. It is obtained by applying both the dc and the ac electric fields. The dc electric field is applied following the procedure described before for the $R_z(-\frac{\pi}{2})$ gate. In addition, when $\Delta E_z(t)=\Delta E_{ct}$ an ac electric field $E_{ac}(t)$ in resonance with the flip-flop transition frequency is applied for a time $T_{E_{ac}}^{ON}$ with $\phi=-\pi/2$. In Table \ref{tab:H} all the parameters set to implement the H gate are reported.
\begin{widetext}
\begin{center}
\begin{table}[htbp!]
\caption{\label{tab:H}H gate parameters}
\begin{center}
\begin{ruledtabular}
\begin{tabular}{cccccccccccc}
$\Delta E_{idle}$ & $\Delta E_{int}$ & $\Delta E_{ct}$ & $E_{ac}$ & $\tau_1$ & $\tau_2$ & T & $T_{E_{ac}}^{ON}$ &  $V_t$ & K & $K_E$ & $T_{gate}$\\
{[V/m]} & {[V/m]} & {[V/m]} & {[V/m]} & {[ns]} & {[ns]} & {[ns]} & {[ns]} &{[GHz]} &  & & {[ns]}\\
\hline
10000 & 500 & 0 & 180 & 1.7 & 3.5 & 41.5 &  40 &  11.5 & $\simeq$21 & $\simeq$57 & 51.9\\
\end{tabular}
\end{ruledtabular}
\end{center}
\end{table}
\end{center}
\end{widetext}

For the H gate, in addition to the adiabaticity factor $K$, we evaluated $K_E$ using $\Delta_E=\pi\delta_E$ where $\delta_E=\omega_E/(2\pi)-\varepsilon_0$ and $\Omega_E= 2\pi g_E$ where $g_E=\frac{edE_{ac}}{4h}\frac{V_t}{\varepsilon_0}$. 

\begin{widetext}

\begin{figure}[htbp!]
\centering
a)\includegraphics[width=0.8\textwidth]{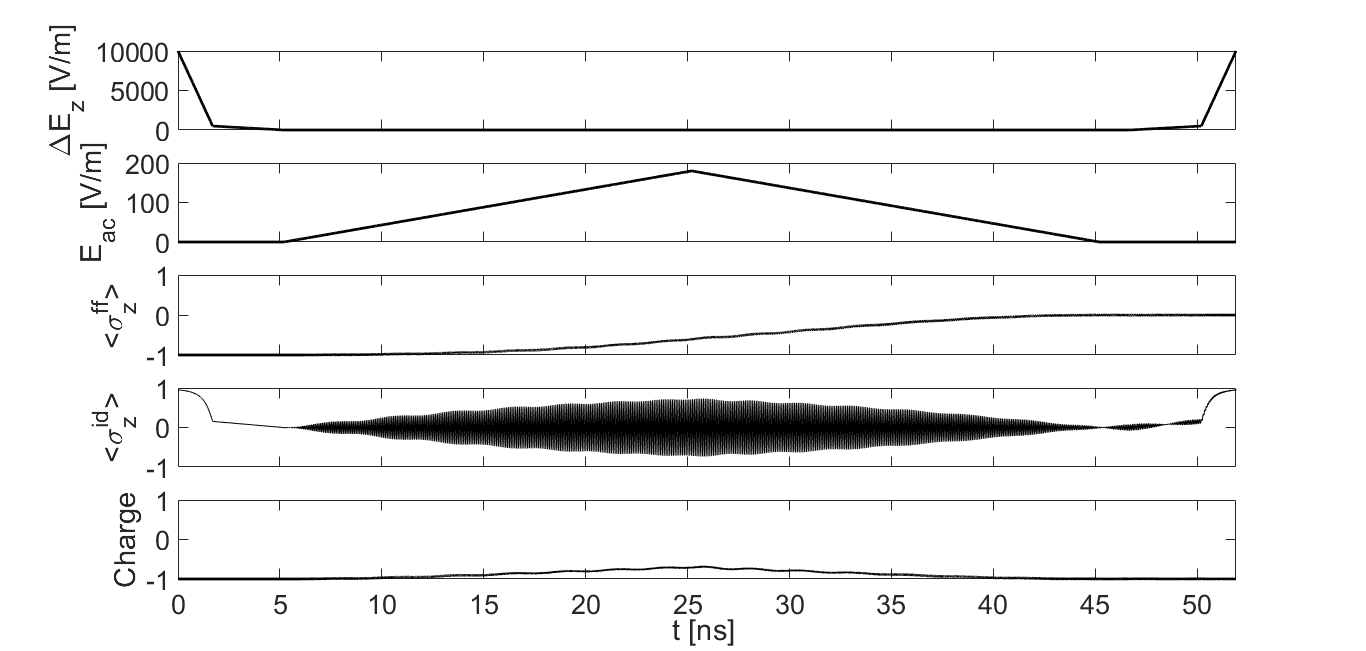}\\ b)\includegraphics[width=0.8\textwidth]{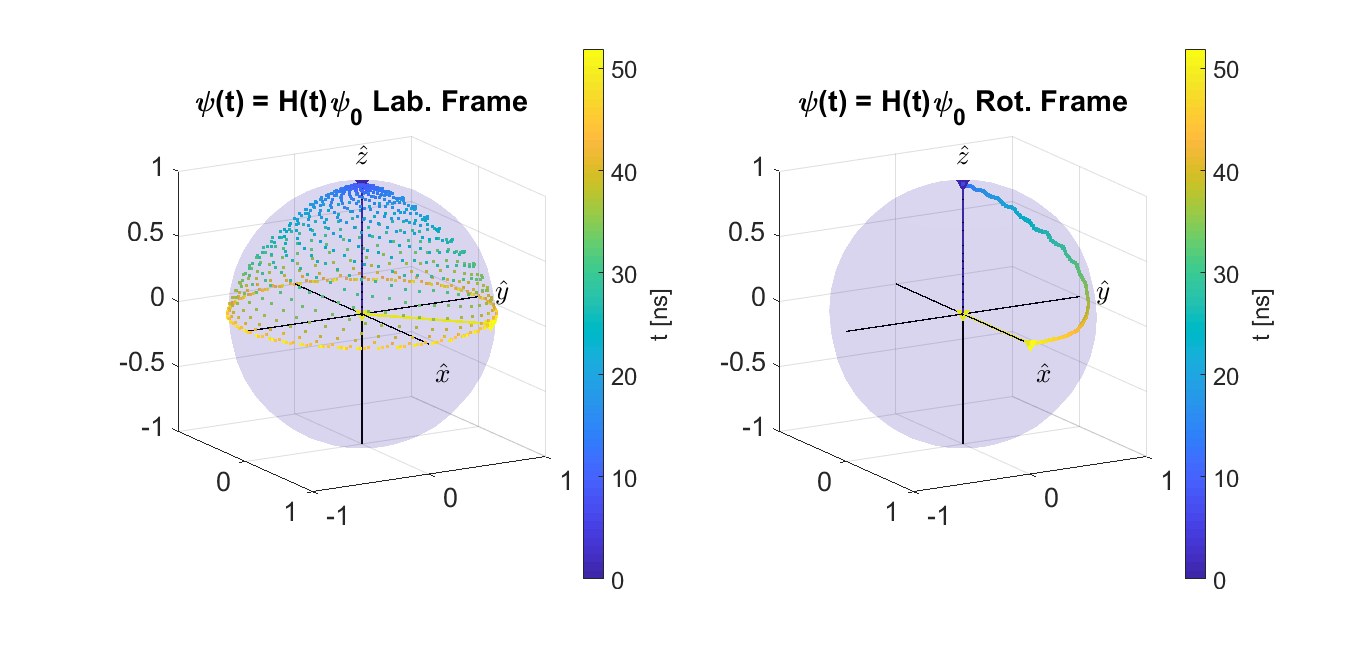}\\
\caption{(a) Time evolution of the dc applied electric field $\Delta E_z(t)$, of the amplitude of the ac electric field $E_{ac}(t)$ and of the mean values of the operators $\sigma^{ff}_z$, $\sigma^{id}_z$ and the charge operator $\ket{e}\bra{e}-\ket{g}\bra{g}$ in the flip-flop subspace during the evolution of the H gate for the initial state $\ket{\psi_0}=\ket{0}$. (b) Bloch sphere representation of the H gate. The blue (yellow) arrow represents the initial (final) state in the laboratory frame (left) and in the rotating frame (right).}\label{H}
\end{figure}

\end{widetext}

Analogously to the $R_z(-\frac{\pi}{2})$ gate, in Fig. \ref{H} we observe the behavior of the H gate starting from the qubit initial condition $\ket{\psi_0}=\ket{0}$.

\subsubsection{Entanglement fidelity}
In order to assess the performance of our gates in the presence of noise, we adopt the entanglement fidelity $F$ \cite{Nielsen-2000,Marinescu-2012}, that does not depend on the qubit initial condition, and is defined as
\begin{equation}
F=tr[\rho^{RS}\mathbb{1}_{R}\otimes(U_{i}^{-1}U_{d})_{S}\rho^{RS}\mathbb{1}_{R}\otimes(U_{d}^{-1}U_{i})_{S}],
\end{equation}
where $U_{d}$ ($U_{i}$) is the disturbed (ideal) quantum gate and $\rho^{RS}=\ket{\psi}\bra{\psi}$ where $\ket{\psi}$ represents a maximally entangled state in a double state space generated by two identical Hilbert spaces $R$ and $S$, that is $\ket{\psi}=\frac{1}{\sqrt{2}}(|00\rangle+|11\rangle$ for the single qubit gates and $\ket{\psi}=\frac{1}{\sqrt{2}}(|0000\rangle+|1111\rangle$ for the two-qubit gate. 

\begin{figure}[H]
\centering
\includegraphics[width=\columnwidth]{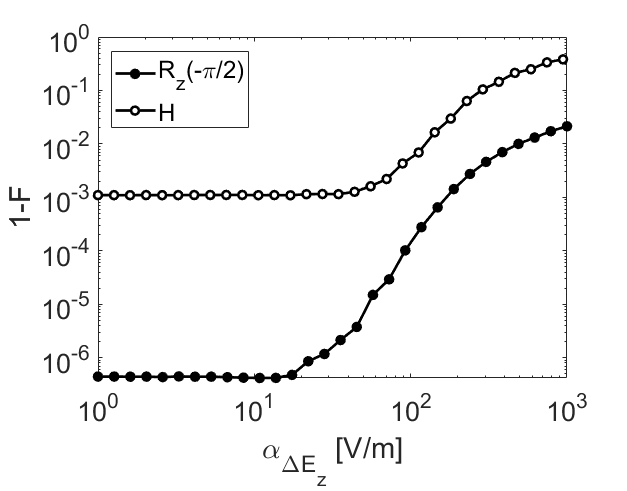}
\caption{Entanglement infidelity for the $R_z(-\frac{\pi}{2})$ and H gates as a function of the noise amplitude $\alpha_{\Delta E_z}$.}\label{Fid}
\end{figure}

In Fig. \ref{Fid} we show the entanglement infidelity 1-F for the $R_z(-\frac{\pi}{2})$ and the H gates when a noise amplitude $\alpha_{\Delta E_z}$ on the electric field $\Delta E_z$ in the interval $[1,1000]$ V/m is considered.

For both the quantum gates we observe the same qualitative behaviour of the infidelity. In the intervals $\alpha_{\Delta E_z}\simeq[1,20]$ V/m for $R_z(-\frac{\pi}{2})$ and $\alpha_{\Delta E_z}\simeq[1,40]$ V/m for H, the infidelities show a plateau that reflects the non-adiabaticity of the sequence. Increasing the value of the coefficient $K$ leads to a more adiabatic operation that returns a lower value of the infidelities in the plateau. In this region, the $R_z(-\frac{\pi}{2})$ gate shows the higher values of fidelity, that is, around 99.9999\%, followed by the H gate fidelity that starts approximately from 99.9\%. Then the infidelities slowly grow up until they settle to higher values in correspondence to high values of the noise amplitude. For all the gates under study, the fidelities show very promising values up to very reasonable values of the experimental noise amplitude, i.e. $\alpha_{\Delta E_z}\leq100$ V/m. Indeed, we have F$\geq$99.99\% for the $R_z(-\frac{\pi}{2})$ gate and F$\geq$99.3\% for the H gate.

\section{Two flip-flop qubits}
The universal set of quantum gates may be completed by the $\sqrt{iSWAP}$ two-qubit gate. In the first part of this Section, we present the Hamiltonian model describing two interacting flip-flop qubits, whereas in the second part the $\sqrt{iSWAP}$ is derived and the effects of the noise are investigated.

\subsection{Hamiltonian model}
The two flip-flop qubits Hamiltonian model $H^{2ff}$ is obtained adding up two single-qubit Hamiltonians, supposed identical, and an interaction term
\begin{equation}
H^{2ff}=H^{ff}\otimes\mathbb{1}+\mathbb{1}\otimes H^{ff}+H_{int}.
\end{equation}
$H_{int}$ is the dipole-dipole interaction and is equal to
\begin{equation}
H_{int}=\frac{1}{4\pi\epsilon_0\epsilon_r r^3}\left[\textbf{p}_1\cdot\textbf{p}_2-\frac{3(\textbf{p}_1\cdot\textbf{r})(\textbf{p}_2\cdot\textbf{r})}{r^2} \right],
\end{equation}
where $\epsilon_0$ $(\epsilon_r)$ is the vacuum permittivity (material dielectric constant, that we set to the silicon value 11.7) and $\textbf{r}$ is the two-qubit distance. The dipole operator is $p_k=de(\mathbb{1}+\sigma_{z,k}^{id})/2$, $(k=1,2)$, and we assume that the dipoles are oriented perpendicularly to their separation, i.e. $\textbf{p}_k\cdot\textbf{r}=0$. From all these considerations, we have
\begin{equation}
H_{int}=\frac{d^2e^2}{16\pi\epsilon_0\epsilon_r h r^3}(\mathbb{1}_1\mathbb{1}_2+\sigma_{z,1}^{id}\mathbb{1}_2+\mathbb{1}_1\sigma_{z,2}^{id}+\sigma_{z,1}^{id}\sigma_{z,2}^{id}).
\end{equation}

\subsection{Two-qubit gate: $\sqrt{iSWAP}$}
The matrix that represents the $\sqrt{iSWAP}$ gate in the two-qubit logical basis is given by
\begin{equation}\label{mswap}
\sqrt{iSWAP}=
\begin{pmatrix}
1 & 0 & 0 & 0 \\
0 & \frac{1}{\sqrt{2}} & \frac{1}{\sqrt{2}}i & 0 \\
0 & \frac{1}{\sqrt{2}}i & \frac{1}{\sqrt{2}} & 0 \\
0 & 0 & 0 & 1
\end{pmatrix}
\end{equation}
and the set of parameters included into the sequence that realize the transformation in Eq.(\ref{mswap}) is reported in Table \ref{tab:swap}. 
\begin{widetext}
\begin{center}
\begin{table}[htbp!]
\caption{\label{tab:swap}$\sqrt{iSWAP}$ gate parameters}
\begin{center}
\begin{ruledtabular}
\begin{tabular}{cccccccccc}
 & $\Delta E_{idle}$ & $\Delta E_{int}$ & $\Delta E_{ct}$ & $\tau_1$ & $\tau_2$ & T & $V_t$ & K & $T_{gate}$\\
 &{[V/m]} & {[V/m]} & {[V/m]} & {[ns]} & {[ns]} & {[ns]} & {[GHz]} & & {[ns]}\\
\hline
$Q_1Q_2$ &10000 & 500 & 0  & 1.7 & 99 & 2 & 11.58 & $\simeq$21 & 203.4\\
\hline
$Q_1 (Q_2)$ &10000 & 500 & 0 & 1.7 & 3.5 & 1.2 & 11.58 & $\simeq$21 & 11.6\\
\end{tabular}
\end{ruledtabular}
\end{center}
\end{table}
\end{center}
\end{widetext}

The operation is obtained by first applying to both the qubits $Q_1Q_2$ a dc electric field $\Delta E_z(t)$ with $\tau_1$= 1.7 ns, $\tau_2$=99 ns and T=2 ns, and then by applying two identical single qubit rotations to $Q_1$ and later to $Q_2$ along the $\hat{z}$ axis with $\tau_1$= 1.7 ns, $\tau_2$=3.5 ns and T=1.2 ns, that corresponds to a rotation angle $\theta\simeq -0.5$ rad. When $Q_1$ performs the $\hat{z}$-rotation, $Q_2$ is in $\Delta E_{idle}$, and viceversa. The total time to perform the $\sqrt{iSWAP}$ is given by $T_{gate}=T_{gate}^{Q_1Q_2}+T_{gate}^{Q_1}+T_{gate}^{Q_2}$= 226.6 ns. 

\begin{widetext}

\begin{figure}[htbp!]
\centering
a)\includegraphics[width=0.8\textwidth]{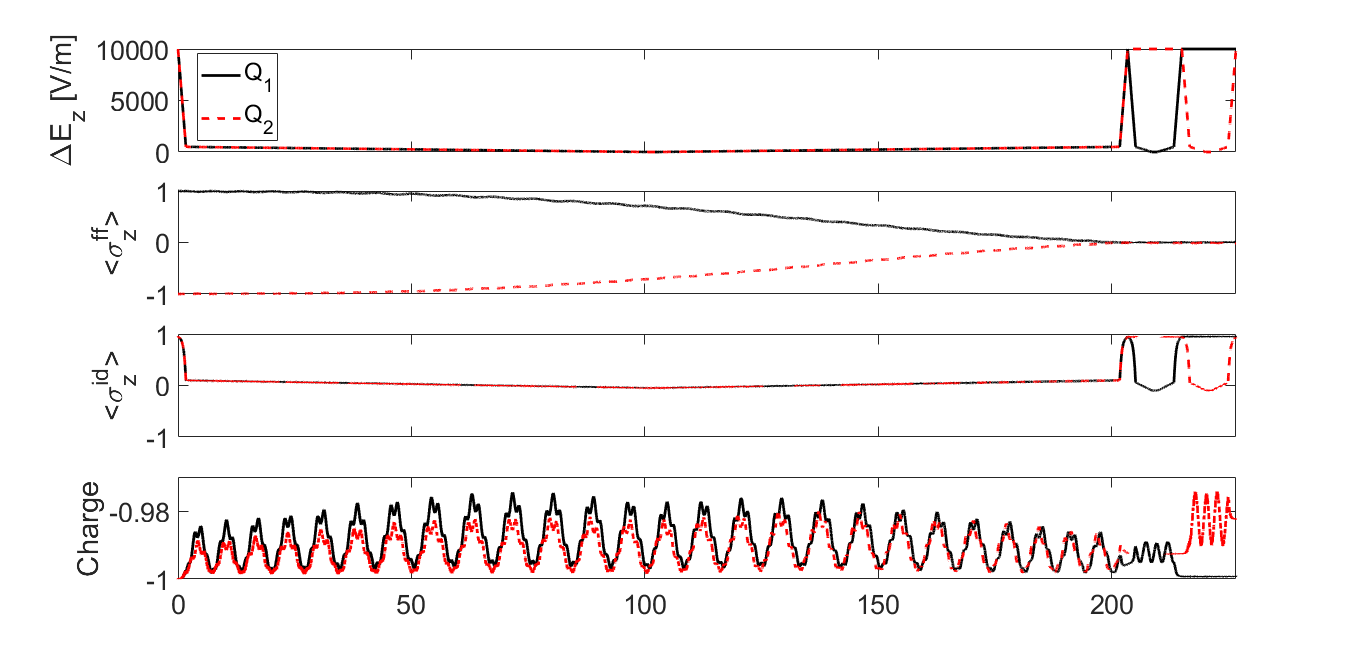}\\ b)\includegraphics[width=0.8\textwidth]{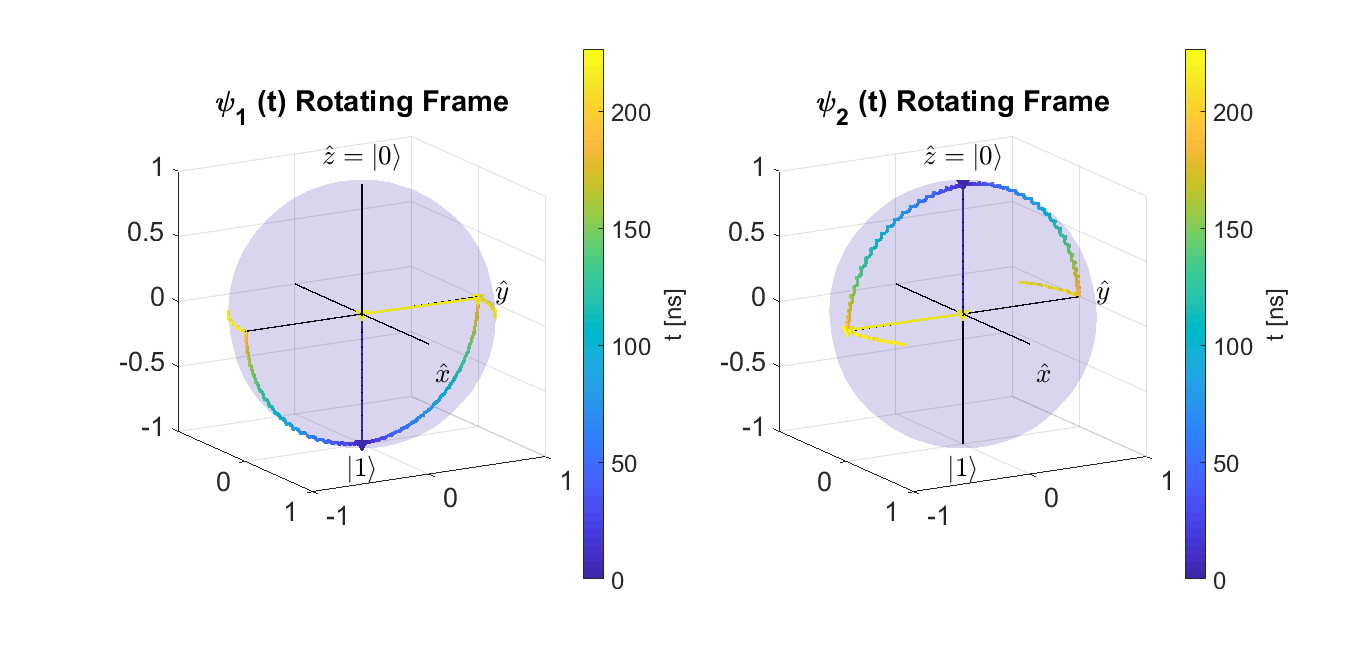}\\
\caption{(a)  Time evolution of the dc applied electric fields $\Delta E_{z,1}(t)$ and $\Delta E_{z,2}(t)$ and of the mean values of the operators $\sigma^{ff}_{z,(1,2)}$, $\sigma^{id}_{z,(1,2)}$ and the charge operator $\ket{e}\bra{e}-\ket{g}\bra{g}$ for the two flip-flop qubits during the evolution of the $\sqrt{iSWAP}$ gate for the initial state $\ket{\psi_{0}}=\ket{\psi_{0_1}}\otimes\ket{\psi_{0_2}}$ with $\ket{\psi_{0_1}}=\ket{1}$ and $\ket{\psi_{0_2}}=\ket{0}$. (b) Bloch sphere representation of the $\sqrt{iSWAP}$ gate. The blue (yellow) arrow represents the initial (final) state for $Q_1$ (left) and $Q_2$ (right) in the rotating frame.}\label{swap}
\end{figure}
\end{widetext}

The dynamical behaviour of the two electric fields $\Delta E_{z,1}(t)$ and $\Delta E_{z,2}(t)$ applied respectively to $Q_1$ and to $Q_2$ are shown in Fig. \ref{swap}(a). In addition, the mean values of the operators for both the qubits are shown. In Fig. \ref{swap}(b) we report the dynamical behaviour on the Bloch sphere during the application of the entire sequence for $Q_1$ (left) and $Q_2$ (right) in the rotating frame, starting from the initial condition $\ket{\psi_{0}}=\ket{\psi_{0_1}}\otimes\ket{\psi_{0_2}}$ with $\ket{\psi_{0_1}}=\ket{1}$ for $Q_1$ and $\ket{\psi_{0_2}}=\ket{0}$ for $Q_2$.

\begin{figure}[htbp!]
\centering
\includegraphics[width=\columnwidth]{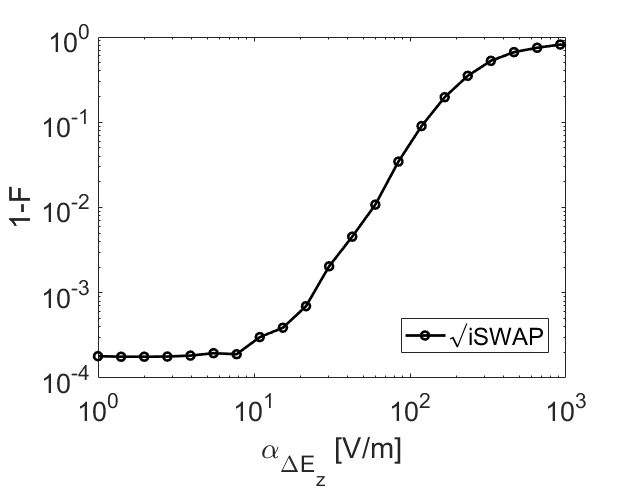}
\caption{Entanglement infidelity for the $\sqrt{iSWAP}$ gate as a function of the noise amplitude $\alpha_{\Delta E_z}$.}\label{Fid2}
\end{figure}

In Fig. \ref{Fid2} we report the entanglement infidelity for the $\sqrt{iSWAP}$ gate when a noise amplitude $\alpha_{\Delta E_z}$ in the interval $[1,1000]$ V/m is considered.

When the noise amplitude lies in the interval $[1,10]$ V/m, the fidelity is F$\simeq$99.98\% and it remains larger than F$\simeq$99.5\% up to $\alpha_{\Delta E_z}\simeq50$ V/m.

\section{Conclusions}
In this paper we have addressed quantum computation by flip-flop qubits, a donor-based qubits in which the logical states are encoded in the donor nuclear and its bound electron. Flip-flop qubits represent an interesting advancement compared to the Kane's seminal proposal, due to the possibility of exploiting the long range electric dipole-dipole interaction. A universal set of quantum gates composed by $\{R_z(-\frac{\pi}{2}),H,\sqrt{iSWAP}\}$ has been presented, and the noise effect on the entanglement fidelity has been studied. The noise model adopted shows a 1/f spectrum, typical of qubits sensitive to charge noise. In terms of fidelity, results are very promising: for example in correspondence to a realistic noise level around 50 V/m, we obtain F$\geq$99.999\% for the $R_z(-\frac{\pi}{2})$ gate and 99.8\% for the H gate. Under the same condition, the two-qubit $\sqrt{iSWAP}$ gate may be realized with a fidelity above 99.5\%. We conclude that flip-flop qubits with long range coupling represent a promising platform for solid state quantum computation.

%

\end{document}